\documentclass[12pt]{iopart}
\usepackage{iopams}
\usepackage{graphicx}
\usepackage{epsfig}
\usepackage{epstopdf}
\usepackage[usenames,dvipsnames]{color}

\newcommand{\be}{\begin{equation}}
\newcommand{\ee}{\end{equation}}
\newcommand{\bea}{\begin{eqnarray}}
\newcommand{\eea}{\end{eqnarray}}

\begin{document}
\title[\small{Thick braneworlds generated by a non-minimally coupled scalar field and
a Gauss-Bonnet term}]{Thick braneworlds generated by a non-minimally coupled scalar field and
a Gauss-Bonnet term: conditions for localization of gravity}

\author{Alfredo Herrera--Aguilar$^{1,2}$, Dagoberto Malag\'on--Morej\'on$^1$, Refugio Rigel Mora--Luna$^1$
and Israel Quiros$^3$}

\address{$^{1}$Instituto de F\'{\i}sica
y Matem\'{a}ticas, Universidad Michoacana de San Nicol\'as de Hidalgo, Edificio C--3, Ciudad
Universitaria, C.P. 58040, Morelia, Michoac\'{a}n, M\'{e}xico}

\address{$^2$Centro de Estudios en F\'{\i}sica y Matem\'{a}ticas B\'{a}sicas y Aplicadas, Universidad
Aut\'{o}noma de Chiapas, Calle 4a Oriente Norte 1428, Tuxtla Guti\'{e}rrez, Chiapas, M\'{e}xico}

\address{$^3$Divisi\'{o}n de Ciencias e Ingenier\'{\i}a de la Universidad de Guanajuato,
A.P. 150, 37150, Le\'{o}n, Guanajuato, M\'{e}xico}

\eads{alfredo.herrera.aguilar@gmail.com, malagon@ifm.umich.mx, rigel@ifm.umich.mx,
iquiros@fisica.ugto.mx}
\date{\today}

\begin{abstract}
We consider warped five-dimensional thick braneworlds with four-dimensional Poincar\'e invariance
originated from bulk scalar matter non-minimally coupled to gravity plus a Gauss-Bonnet term. The
background field equations as well as the perturbed equations are investigated. A relationship
between 4D and 5D Planck masses is studied in general terms. By imposing finiteness of the 4D
Planck mass and regularity of the geometry, the localization properties of the tensor modes of the
perturbed geometry are analyzed to first order, for a wide class of solutions. In order to explore
the gravity localization properties for this model, the normalizability condition for the lowest
level of the tensor fluctuations is analyzed. It is found that for the examined class of
solutions, gravity in 4 dimensions is recovered if the curvature invariants are regular and the 4D
Planck mass is finite. It turns out that both the addition of the Gauss-Bonnet term and the
non-minimal coupling between the scalar field and gravity {\it reduce} the value of the 4D Planck
mass compared to its value when the scalar field and gravity are minimally coupled and the
Gauss-Bonnet term is absent. The above discussed analysis depends on the explicit form of the
scalar field (through its non-minimal coupling to gravity), making necessary the construction of
explicit solutions in order to get results in closed form, and is illustrated with some examples
which constitute smooth generalizations of the so-called Randall-Sundrum braneworld model. These
solutions were obtained by making use of a detailed {\it singular perturbation theory} procedure
with respect to the non-minimal coupling parameter between the scalar field and gravity, a
difficult task that we managed to perform in such a way that all the physically meaningful
conditions for the localization of gravity are fully satisfied. From the obtained explicit
solutions we found an interesting effect: when we consider a non-minimally coupled scalar-tensor
theory, there arise solutions for which the symmetries of the background geometry are not
preserved by the scalar matter energy density distribution. In particular, the value of the ``5D
cosmological constant'' of the asymptotically $AdS_5$ space-time (which is even with respect to
the extra coordinate) gets different contributions at $-\infty$ and $+\infty$ from the asymptotic
values of the self-interaction potential of the scalar field. Thus, an asymmetric energy density
distribution of scalar matter gives rise to a completely even with respect to the fifth coordinate
space-time, in contrast to braneworld models derived from minimally coupled scalar-tensor
theories, where both entities possess the same symmetry.
\end{abstract}
\pacs{11.25.Mj, 04.40.Nr}

\section{Introduction}

Since little more than one decade an interesting alternative to the standard Kaluza-Klein
compactification paradigm has been put to test. The kind of scenarios based on this alternative
are known as braneworld models \cite{anto1}--\cite{tanaka} and allow for infinite extra
dimensions, in contrast to the Kaluza-Klein idea, where extra dimensions are compactified to a
very small size. The mentioned alternative requires the standard model (SM) fields to be trapped
on a 4D hypersurface, called a 3-brane. Unlike ordinary matter, gravitons and exotic matter are
allowed to propagate in the bulk of the higher-dimensional manifold. Since gravity can propagate
through all dimensions, the first important question concerning to braneworld models is to check
whether they give back standard 4D gravity on the brane.

In the thin braneworld models version \cite{merab}--\cite{rs2} the branes are modeled by 4D delta
functions, a mathematical complication that requires the fulfillment of the so-called
Israel-Lanczos junction conditions, which, basically, dictate the way the brane must be embedded
into the higher-dimensional bulk to accommodate the matter degrees of freedom (SM particles) that
are trapped on it. These junction conditions become much more complicated when adding to the setup
higher curvature terms, for instance \cite{junctionconds}. Moreover, in these models the curvature
is singular at the location of the branes, a drawback from the gravitational point of view that
can be healed in several ways.

On the other hand, thin brane models are just an idealization of the physical reality.
Braneworlds, if they are to be considered as models for our world, have to be of finite thickness.
Actually, at high enough energies, the SM particles might acquire a small (but non-negligible)
momentum in the extra space. Indeed, the original braneworld idea put forth in reference
\cite{arkani} is consistent with a non-vanishing brane thickness $\sim m^{-1}_{EW}$ ($m_{EW}$ is
the electro-weak energy scale).\footnote{Electroweak interactions have been probed, precisely, at
distances $m^{-1}_{EW}$.} These more realistic alternatives to thin brane configurations are
known, generically, as thick braneworlds, or, also, domain walls.

Thick branes might be generated in a variety of ways. One example is by replacing the delta
functions in the action either by other distribution functions (see, for instance references
\cite{csaba} and \cite{soda}), or, alternatively, by self-interacting scalar fields minimally
coupled to gravity \cite{gremm}--\cite{barbosa1}. Within the framework of thin (Randall-Sundrum)
braneworlds, bulk scalar fields have been investigated, for instance, in
\cite{farakos,farakos2,ss}, where the Einstein-Hilbert action for Randall-Sundrum branes has been
modified by considering self-interacting scalar fields non-minimally coupled to gravity. In
\cite{farakos2} the resulting action has been further modified by adding a Gauss-Bonnet term,
whereas cosmological applications have been considered in \cite{ss}. Possible influence of higher
curvature terms in scalar-field-generated brane models has been studied in references
\cite{mg}--\cite{corradini}. The thick braneworld approaches avoid solving the Israel-Lanczos
junction conditions always present when studying  braneworlds modeled by 4D delta functions (with
or without the presence of scalar fields). Instead, a relevant differential equation must be
solved either for the smooth distribution function or the scalar field of the setup. This task is,
however, easier than facing the mathematical difficulties of solving the junction conditions when
considering higher curvature terms in the setup.

In dimensions larger than four (in 5D, for instance) the usual Einstein-Hilbert action may be
supplemented with higher order curvature corrections. For some special cases, these corrections
lead to equations of motion with at most second order derivatives of the metric with respect to
space-time coordinates \cite{lovelock-madore}. A particular combination containing higher
curvature terms which yields second order differential equations is the well-known Gauss-Bonnet
(GB) invariant:

\be {\cal R}^2_{\rm GB}=R^{ABCD} R_{ABCD}-4R^{AB} R_{AB}+R^2,\label{GB-term}\ee where $A, B, C, D
= 0,1,2,3,5$. While in four dimensions the GB invariant is a topological term -- in other words,
it can be arranged into a four-divergence -- which does not contribute to the classical equations
of motion, in more than 4D the GB combination leads to a theory free of spin 2 ghosts due to
higher derivatives and it appears in different higher dimensional contexts. For instance, in
string theory, the heterotic string $\alpha$ correction is fixed to be the Gauss-Bonnet term in
order to avoid a ghost to that order in the (tree-level) effective action
\cite{GB-string}--\cite{grosssloan}.\footnote{One finds a Riemann square term in the first string
tension correction and the remaining terms are added by hand by field redefinitions that do not
affect the effective action to that order. However, it is worth noticing that if one does not
choose the GB combination of the squared curvature terms, the corresponding graviton propagator
would have terms with an arbitrary number of derivatives \cite{grosssloan}.} However, in a second
order derivative theory there is a sector that might contain spin 2 ghosts that are not originated
by higher derivatives. If we wish to study braneworlds free of such ghosts, we need to carefully
monitor the overall sign of the norm of the graviton spectrum. Within the context of thick
braneworld scenarios generated by a self-interacting minimally coupled scalar field, the GB
invariant has been studied in connection with the localization properties of the various modes of
the geometry in Ref. \cite{mg} (see also, for instance, \cite{corradini}). Non-minimal interaction
of the bulk (self-interacting) scalar field has been considered, for instance, in Refs.
\cite{andrianov}, where the perturbative stability of the configurations was explored, and also
within the framework of Weyl geometry in \cite{ariasetal}--\cite{bh3}.

It seems quite natural to further explore the influence of
non-minimal coupling of a self-interacting scalar field with
gravity, within the framework of thick braneworlds with higher
curvature terms (a Gauss-Bonnet invariant, for instance). In this
work, we aim, precisely, at studying a 5D thick braneworld modeled
by a smooth scalar domain wall non-minimally coupled to gravity with
a Gauss-Bonnet term on the bulk. To be precise we shall consider
specifically conformally flat geometries. We shall focus here in the
investigation of the relationship among: localization properties of
the tensor zero mode, finiteness of 4D Planck mass and smoothness of
the geometry, for a wide class of solutions. This will allow us to
extend and generalize, in particular, previous results obtained
within the framework of braneworlds with 4D Poincar\'e symmetry
generated by scalar fields minimally coupled to gravity \cite{hmmn}
(see also \cite{liuetal} for a similar approach within braneworlds
with de Sitter 3-branes). In order to do that, we analyze the tensor
perturbations of the geometry.

In general terms, the study of metric fluctuations in braneworld
models is complicated because of the coupling between geometry and
matter at the field equations level. On the one hand, the
perturbations problem is more tractable if the background metric has
some isometries. In this case, the metric perturbations can be
classified according to the symmetry of the problem. If the metric
respects 4D Poincar\'e symmetry, then, the different perturbation
modes of the geometry can be classified in a gauge-invariant manner
into scalar, vector an tensor modes under 4D Poincar\'e
transformations. Furthermore, in \cite{mg1,andrianov} it was shown
in detail that each fluctuation mode evolves independently, then, if
one studies the evolution of a specific mode it is not necessary to
worry about a possible coupling with the remaining perturbation
modes. On the other hand, in order to recover the standard 4D
gravity predicted by General Relativity, the existence of a 4D
massless spin 2 field localized on the brane is necessary. If there
is a 4D massless spin 2 field in our model it should be described by
the propagating tensor modes of our 5D perturbation problem. Not
only the gravitational perturbation equations, but also the
background equations are complicated and it is very difficult to
analytically solve them. In spite of this difficulty, here we solve
these equations for a non-singular geometry. We consider the effects
of the non-minimal coupling of the bulk scalar field with the
curvature as a small perturbation characterized by a small
dimensionless parameter. Usually, in this kind of problems there are
regions -- called boundary layer regions \cite{eckhaus}) -- where
the formal expansion with respect to the small parameter is not
valid. Then, the expansion must be redefined for these regions.
After having all the asymptotic expansions on the different regions,
it is necessary to combine them to get an expansion valid on the
entire domain.

The paper has been organized as follows. In Sec. II we present the
model setup and the corresponding field equations -- for the
respective metric ansatz -- are given. In Sec. III it is further
shown that 4D gravity can be localized on this particular braneworld
and the normalization condition for a massless spin 2 fluctuation
mode is derived. The relationship between the 4D and 5D Planck
masses is obtained in Sec. IV, by integrating the action of the
model with respect to the fifth dimension. In Sec. V a class of
solutions for which the geometry is singularity-free is presented.
Then, in Sec. VI, we further construct some approximate analytical
solutions for the scalar field after fixing the form of the warp
factor, as well as the functional dependence of the non-minimal
coupling of the scalar field to gravity. Three different cases are
considered by appropriate choices of the free parameters. In the
particular case when the Gauss-Bonnet term is switched off, an exact
solution is obtained. Finally, in Sec. VII we present a brief
summary of our results, and conclusions are given.


\section{The model}

Here we shall explore a thick braneworld described by the following
5D action (compare, for instance, with the actions investigated in
references \cite{mg} and \cite{andrianov})
\begin{eqnarray}
&S = \int d^5 x \sqrt{|g|} \biggl\{-\frac{L(\varphi)R}{2\kappa}
-\alpha {\cal R}^2_{\rm GB} + \frac{1}{2} (\nabla \varphi)^2 -
V(\varphi) \biggr\}, \label{action}\end{eqnarray} where $\alpha>0$
(we choose the sign of the $\alpha$ coupling to be positive, although
it could have any sign).
On the other hand $\kappa \simeq 1/M^3$, where $M$ is the
5D Planck mass, the function  $L(\varphi)$ is the coupling between
the scalar field $\varphi$ and the curvature (gravity), while
$V(\varphi)$ is the scalar field's self-interaction potential, and
${\cal R}^2_{\rm GB}$ is the 5D Gauss-Bonnet term defined in Eq.
(\ref{GB-term}). The Einstein's field equations that are derivable
from the action (\ref{action}) are the following:
\begin{eqnarray}
&L\, R_{A B}= \kappa \,\tau_{A B}\,+\nabla_A \nabla_B L +
\frac{1}{3}\,g_{A B}\,\Box L -\epsilon \,{\cal Q} _{A B},
\label{ee}\end{eqnarray} where $\epsilon=2\alpha\kappa$ and
$\Box=g^{CD}\nabla_C \nabla_D.$ The reduced energy-momentum tensor
$\tau_{A B}$ corresponding to the scalar matter content on the bulk
takes the form
\begin{equation}
\tau_{A B} = \partial_{A} \varphi \partial_{B} \varphi -
\frac{2}{3}\, g_{A B} \,V(\varphi).\nonumber\end{equation} The term
${\cal Q}_{A B}$ is called Lanczos tensor  and represents the
corrections of the Gauss-Bonnet term to  the Einstein equations,
which can be written in the form
\begin{eqnarray}
\!\!\!\!\!\!\!\!\!\!\!\!\!\!\!\!\!\!\!\!\!\!\!\!\!\!\!\!&{\cal Q}_{A B} =\frac{1}{3} \,g_{A B}\, {\cal R}^2_{\rm GB} -
2\,R\, R_{A \,B} + 4\,R_{A\,C} \, R^{C}\,_{B} + 4
R^{C\,D}\,R_{A\,C\,B\,D}-2\,R_{A\,C\,D\,E}\,
R_{B}\,^{\,C\,D\,E}.\nonumber\end{eqnarray} Finally, the remaining
terms in the right hand side of (\ref{ee}) come from the non-minimal
coupling of the scalar field to gravity.

The Klein-Gordon equation determining the dynamics of the scalar
field is obtained by varying the action (\ref{action}) with respect
to $\varphi$:
\begin{eqnarray}
\Box\varphi+\frac{1}{2\kappa}\,R\,L_{\varphi}+\frac{\partial
V}{\partial\varphi} =0,\quad \mbox{where} \quad
L_{\varphi}=\frac{dL}{d \varphi}. \label{kg}\end{eqnarray}
As in \cite{mg1} let us consider a warped metric in conformally flat coordinates
\begin{equation}
ds^2 = a^2(w) \left[\eta_{\mu \nu}dx^{\mu}dx^{\nu}- dw^2\right],
\label{metric}\end{equation} where the variable $w$ is the extra
coordinate and $\eta_{\mu \nu}$ is the 4D Minkowski metric. Here,
for simplicity, we focus in the case where the scalar field depends
only on the extra-coordinate $w$.

In terms of the above metric ansatz the field equations (\ref{ee})
and (\ref{kg}) read
\begin{eqnarray}
\!\!\!\!\!\!\!\!\!\!\!\!\!\!\!\!\!\!\!\!\!\!\!\!\!&&V  +\frac{3 \,{\cal H}\,L_{\varphi} \,\varphi'}{\kappa
\,a^2}+\frac{1}{2 \kappa \,a^2} \biggl(\varphi''\,L_{\varphi}
+\varphi'^2\,L_{\varphi \varphi}\biggr)- \frac{ 3}{ 2 \kappa
a^2}\biggl[\frac{4\,\epsilon}{a^2}\,{\cal H}^2\,\left({\cal H}^2 +
{\cal H}'\right)-\left({\cal H}'+3{\cal H}^2\right)\,L\biggr]= 0, \nonumber \\
\!\!\!\!\!\!\!\!\!\!\!\!\!\!\!\!\!\!\!\!\!\!\!\!\!\!&&\frac{1}{\kappa} L_{\varphi}\,
\varphi''-\frac{2}{\kappa}\,{\cal H}\,L_{\varphi}\,\varphi'+
\left( 1+\,\frac{L_{\varphi \varphi}}{\kappa}\right)\varphi'^2 -
\frac{3}{\kappa}\left({\cal H}^2-{\cal H}' \right)q = 0, \label{eequations} \\
\!\!\!\!\!\!\!\!\!\!\!\!\!\!\!\!\!\!\!\!\!\!\!\!\!\!&&\varphi''+3{\cal H}\varphi'-\frac{dV}{d\varphi}
a^2-\frac{2L_{\varphi}}{\kappa} \left( 3{\cal H}^2+2{\cal
H}'\right)=0,\nonumber\end{eqnarray} respectively. Here the tilde
denotes derivative with respect to the extra-coordinate
($'=\frac{d}{dw}$), while ${\cal H} = \frac{a'}{a}$, and
$q=L-\frac{4 \, \epsilon}{a^2}\,{\cal H}^2.$

The above are not independent equations. As one can easily check, if
one combines the first two equations, the last one is obtained.

\section{ Localization of gravity}

As it has been mentioned above, in linear perturbation theory there
is no coupling between the different fluctuation modes, and hence,
they evolve independently. In order to investigate if there is a 4D
massless spin 2 field localized on the brane, it suffices to
consider only the linear tensor fluctuations of the background
metric (\ref{metric})
\begin{equation}
\!\!\!\!\!\!\!\!\!\!\!\!\!ds_{p}^2 =\left[a^2(w) \eta_{A B}+H_{AB}\right]dx^{A}dx^{B},
\quad\mbox{where} \quad H_{A B}=a^2(w)
\left( \begin{array}{cc} 2 h_{\mu\nu} & 0\\  0 & 0\\
\end{array}\right).\nonumber\end{equation} Thus, the  tensor
$h_{\mu\nu}$ is a divergence-less and trace-less rank-two tensor
with respect to 4D Poincar\'e transformations \cite{csaba,mg1}.
After taking this fact into account, perturbing equations (\ref{ee})
and (\ref{kg}), and performing some tedious algebra, the tensor
fluctuation equation reads
\begin{equation}
q {h_{\mu}^{\nu}}'' + ( 3{\cal H} q + q' ) {h_{\mu}^{\nu}}' -
\biggl[ q + \frac{(q-L)'}{2{\cal H}}\biggr] \Box^{\eta}
h_{\mu}^{\nu} =0,\label{graviton}
\end{equation} where $\Box^{\eta}$ is the 4D (Minkowski) D'Alambertian.
If we redefine the field $ h_{\mu \nu}$ as $\Psi_{\mu \nu}
=\sqrt{s(w)} h_{\mu \nu}$ the equation (\ref{graviton}) transforms
into
\begin{equation}
\Psi_{\mu \nu}'' - \frac{({\sqrt{s}})''}{\sqrt{s}} \Psi_{\mu \nu} -
\frac{r}{s} \Box^{\eta}\Psi_{\mu
\nu}=0,\label{fluctuations}\end{equation} where $s(w) = a^{3} q$,
and  $r(w) =a^{3}\biggl( q + \frac{(q-L)'}{2{\cal H}}\biggr)$. In
order to explore the mass spectrum of the metric fluctuations we
will assume separation of variables: $\Psi_{\mu
\nu}=\psi(w)\chi_{\mu \nu}(x)$, where the field $\chi_{\mu \nu}(x)$
describes the 4D massive tensorial modes. Thus, equation
(\ref{fluctuations}) splits into two equations:
\begin{eqnarray}
\Box^{\eta} \chi_{\mu \nu}+m^2 \chi_{\mu \nu}&=&0,\label{graviton4d}\\
\psi'' - \frac{({\sqrt{s}})''}{\sqrt{s}} \psi+m^2
\frac{r}{s}\psi&=&0.\label{graviton5d}\end{eqnarray} The function
$\psi(w)$ is the $w-$dependent amplitude of the dynamical field
$\chi_{\mu \nu}(x)$ and defines the localization properties of the
5D field $\Psi_{\mu \nu}$. The tensor zero mode of $\chi_{\mu
\nu}(x)$  can be identified with the 4D massless spin 2 field. This
mode is localized on the brane if the associated zero mode
fluctuation wave function  $\psi_0(w)$ is normalizable, in other
words, the norm for $\psi_0(w)$ has to be finite
\begin{equation}
\langle \psi_0 |\psi_0 \rangle=\int_{-\infty} ^{\infty}
\frac{r}{s}\,\psi_0^2 \,dw < \infty. \label{conditionnorma}
\end{equation}
The massless eigenstate of (\ref{graviton5d}) is
$\psi_{0}=\sqrt{s}$, then the above normalization condition
transforms into:
\begin{eqnarray}
&\langle \psi_0 |\psi_0 \rangle=\int_{-\infty}^{\infty} a^3\,q \,dw
-4 \,\epsilon \,[a']_{-\infty}^{\infty} +8\,\epsilon
\int_{-\infty}^{\infty}
\frac{a'^2}{a}\,dw.\label{norma1}\end{eqnarray}
It is interesting to ask: what is the relationship between the 4D
Planck mass and the normalizability condition
(\ref{conditionnorma})? In the next sections we will answer this
question.

\section{Planck masses}

Consider a space-time of the following form:
\begin{equation}
ds^2 =a^2(w) [\tilde{g}_{\mu \nu}(x)dx^{\mu}dx^{\nu}- dw^2],
\label{stconformal}
\end{equation} where $ \tilde{g}_{\mu \nu}(x)$ is
an arbitrary 4D metric. Then, the relationship between 4D and 5D
Planck masses can be obtained if we perform a dimensional reduction
by integrating (\ref{action}) with respect to the $w$ coordinate.
Since the coupling function $L(\varphi)$ depends only on the extra
coordinate $w$, the 5D theory (\ref{action}) is reduced to a 4D
Einstein-Hilbert effective action plus the corrections that come
from the scalar matter and higher curvature terms of the bulk
\begin{equation}
S_{4}  \simeq M_{\rm Pl}^2\int d^{4} x \sqrt{|\tilde{g}_4|}
\tilde{R}_{4} + \cdots,\nonumber\end{equation} where the subscript
$4$ labels quantities computed with respect to 4D metric
$\tilde{g}_{\mu \nu}(x)$.

A way to find the Einstein-Hilbert part of the 4D effective action
consists in considering the space-time (\ref{stconformal}) as a
conformal transformation of the metric $\tilde{g}_{A B}$ as follows
\cite{faraoni}:

\[
 \tilde{g}_{A B} \rightarrow g_{A B}= a^2(w) \tilde{g}_{ A B}= a^2(w)
 \left( \begin{array}{cc}
\tilde{g}_{\mu \nu}(x) & 0  \\
0 & -1 \\
\end{array} \right),
\] and, rewriting the action (\ref{action}) in terms of the quantities
defined with respect to $\tilde{g}_{A B}$. Since we only need to
find the Einstein-Hilbert part of the 4D effective action, it is
only necessary to apply the above conformal transformation to the
terms $\frac{-L(\varphi)R}{2 \kappa}$ and $\alpha {\cal R}^2_{\rm
GB}$ in the 5D action. Therefore, the following expressions display
the terms we need
\begin{equation}
 R= a^{-2} \left( \tilde{R}-8 \tilde{\Box} \vartheta -12 (\tilde{\nabla}
 \vartheta)^{2}\right),\end{equation}
\begin{eqnarray}\nonumber
\!\!\!\!\!\!\!\!\!\!\!\!\!\!{\cal R}_{GB}^2 &=&\tilde{C}^2 + \frac{1}{3 a^{4}}
\left[-8\tilde{R}_{A B}\tilde{R}^{A B}
+\frac{5}{2}\tilde{R}^{2}-12\tilde{R} (\tilde{\nabla} \vartheta)^2
-24 \tilde{R} \tilde{\Box}\vartheta
+48 \tilde{R}^{A B}\tilde{\nabla}_{A}\tilde{\nabla}_{B}\vartheta \right.\nonumber \\
\!\!\!\!\!\!\!\!\!\!\!\!\!\!&-&48 \tilde{R}^{A B}\tilde{\nabla}_{A} \vartheta \tilde{\nabla}_{B} \vartheta +
72 (\tilde{\Box} \vartheta)^2 +72 (\tilde{\nabla} \vartheta)^4 -
72 (\tilde{\nabla}_{A}\tilde{\nabla}_{B}\vartheta )(\tilde{\nabla}^{A}\tilde{\nabla}^{B}
\vartheta)\nonumber \\
\!\!\!\!\!\!\!\!\!\!\!\!\!\!&+&\left. 144
(\tilde{\nabla}_{A}\tilde{\nabla}_{B}\vartheta)(\tilde{\nabla}^{A}\vartheta)
(\tilde{\nabla}^{B}
\vartheta) +144 (\tilde{\nabla} \vartheta)^2 \tilde{\Box}
\vartheta\right],\nonumber\end{eqnarray}
where $\vartheta=\ln a$.
The quantities with a tilde are defined with respect to metric
$\tilde{g}_{A B}$ and  $\tilde{C}^2 =\tilde{C}^{ A B C D}
\tilde{C}_{A B C D}$, where $\tilde{C}_{A B C D}$ is the Weyl tensor
computed with the metric $\tilde{g}_{A B}$.

After separating the 4D and 5D contributions  of the above
quantities, substituting them in (\ref{action}) and integrating with
respect to the $w$ coordinate, we obtain  the following expression
for $M_{\rm Pl}$
\begin{eqnarray}
\!\!\!\!\!\!\!\!\!\!\!\!\!\!M_{\rm Pl}^2 &\simeq& M^3 \int_{-\infty}^{\infty} a^3(w)\biggl[
L(\varphi) + \frac{4\epsilon}{a^2}({\cal H}^2 + 2{\cal H}') \biggr]
dw \\ \nonumber
\!\!\!\!\!\!\!\!\!\!\!\!\!\!&=&M^3 \int_{-\infty}^{\infty} a^3(w)\,q \,dw + 8 \,M^3 \,\epsilon
\,[a']_{-\infty}^{\infty}.\label{masaplanck}\end{eqnarray} As one
can see, the Planck mass $M_{\rm Pl}$ is closely related to $\langle
\psi_0 |\psi_0 \rangle$, but an extra term $\int_{-\infty}^{\infty}
\frac{a'^2}{a}\,dw$ arises in (\ref{norma1}). A clearer relation
between  $M_{\rm Pl}$ and $\langle \psi_0 |\psi_0 \rangle$ can be
obtained if we analyze the smoothness conditions for the curvature
invariants.

\section{Smoothness of geometry}

A realistic thick braneworld model should not have singularities in
the geometry. Since we are considering thick
braneworlds, the corresponding warp factors, and hence the geometric
invariants of the theory, must be smooth at the position of the
branes. However, for some kind of solutions naked singularities can
develop at the boundaries of the manifold \cite{hmmn, liuetal} and
we should take care of the behaviour of the curvature invariants at
spatial infinity. In consequence, let us assume that the curvature
invariants and warp factor $a$ are regular in the whole space-time
and consider the class of solutions where asymptotically
\begin{eqnarray}
a(w\rightarrow\infty)\simeq
\frac{1}{w^{\gamma}},\label{asint}\end{eqnarray} with $\gamma$ being
a positive constant. For the metric (\ref{metric}) the curvature
invariants are:
\begin{eqnarray}
R &=&\frac{4}{a^2}\biggl( 2 {\cal H}' + 3 {\cal H}^2\biggr),\nonumber\\
R^{A B} R_{A B}&=& \frac{4}{a^4} \biggl( 5 {{\cal H}'}^2+ 9 {\cal H}^4 +
6 {\cal H}' {\cal H}^2 \biggr),\nonumber\\
R^{ A B C D} R_{A B C D} &=& \frac{4}{a^4} \biggl( 4 {{\cal H}'}^2 + 6{\cal H}^4
\biggr).\nonumber\end{eqnarray} By calculating the asymptotic behaviour of the previous
expressions at $w\rightarrow\infty$ we get: $$ R \simeq w^{2(\gamma -1)},\,\,\,\, R^{A B} R_{A B}
\simeq R^{ A B C D} R_{A B C D}\simeq w^{4(\gamma -1)}.$$ Thus, by imposing smoothness at
infinity, the values of the constant $\gamma$ are restricted to the interval $0<\gamma \leq 1.$
Under this restriction for $\gamma$, we obtain
\begin{eqnarray}
\!\!\!\!\!\!\!\!\!\!&M^2_{\rm Pl}\simeq M^3 \int_{w_{\infty}}^{\infty} a^3\,L
\,dw+\cdots \quad\quad \mbox{and} \quad \quad \langle \psi_0|\psi_0
\rangle\simeq \int_{w_{\infty}}^{\infty} a^3\,L
\,dw+\cdots,\label{finito}\end{eqnarray} where $\cdots$ denote
finite terms and the interval $(w_{\infty},\infty)$ formally
represents the range where the approximation (\ref{asint}) is valid.
Then, for this class of solutions where the geometry is regular, a
finite 4D Planck mass implies localization of gravity on the brane.

In order to have a physically consistent model it is not enough to
satisfy the above conditions. Since, some terms in relations
(\ref{norma1}), (\ref{masaplanck}) and in the definition of
$L(\varphi)$ itself have contributions with negative sign, then, it
is additionally required to check whether the following conditions
are satisfied:

\be L(\varphi)>0,\;\;\langle\psi |\psi\rangle>0,\;\;M_{\rm Pl}^2>0,
\label{ligaduras}\ee where a positive value of the coupling
$L(\varphi)$ is required in order to have a positive definite
quadratic Hamiltonian for the metric tensor modes \cite{andrianov},
while the positive norm of the graviton fluctuation modes is required
in order to have a theory free of spin 2 ghosts.

\section{Some particular solutions}

\begin{figure}[htb]
\begin{center}
\includegraphics[width=8cm]{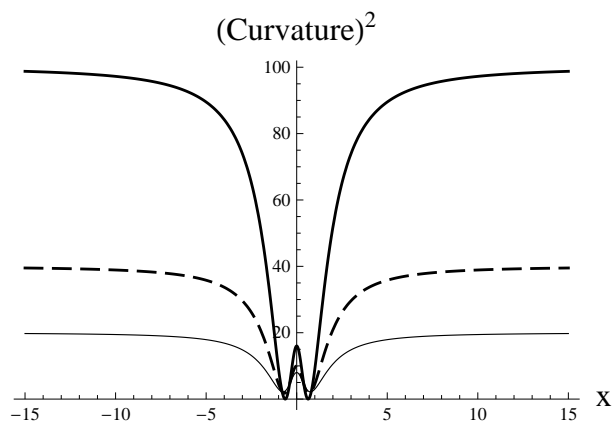}
\end{center}
\caption{Behavior of the different curvature invariants with respect
to the extra space ($a_{0}=b=1$). The thick line represents the
behaviour of $R^2$, the dashed line corresponds to $R^{A B} R_{A B}$
as function of $x$, while the thin line represents the behaviour of
the Kretschmann scalar $ R^{ A B C D} R_{A B C D}$.}
\label{curvaturas}\end{figure}

Let us consider the case where the coupling between the scalar field
and gravity takes the form \cite{farakos} (see also
\cite{andrianov})
\begin{equation}
L=1-\frac{\xi}{2} \varphi^2.\label{pbd}\end{equation} If the
parameter $\xi=0$ the model reduces just to the one explored in Ref.
\cite{mg} (the bulk scalar field is minimally coupled to the
curvature). Hence, the parameter $\xi$ switches between models with
minimal and non-minimal couplings respectively. On the other hand,
we consider a regular metric that interpolates between two
asymptotically $AdS_5$ \cite{mg} space-times. This geometry is
described by the following warp factor:
\begin{equation}
a=\frac{a_0}{\sqrt{1+(bw)^2}}, \label{space-time}\end{equation}
where $\frac{1}{b}$ characterizes the width of the thick brane and
the parameter $a_0$ is related to the radius of the asymptotic
$AdS_5$ space. For this geometry all of the quadratic curvature
invariants are regular and asymptotically constant as it is shown in
Fig. \ref{curvaturas}.

The scalar field $\varphi$ and the self-interaction potential
$V(\varphi)$ can be determined using the first two equations of the
system (\ref{eequations}), since only two equations are independent.
Let us to replace the variable $w$ by the dimensionless variable
$x=bw$. In terms of this new variable the second equation in
(\ref{eequations}) can be rewritten as:
\begin{equation}
\!\!\!\!\!\!\!\!\!\!\!\!\!\!\!\!\!\!\!\!\!\!\!\!\!\!\!
\xi \varphi \varphi'' +\frac{2 \xi x}{1+x^2} \varphi \varphi'
+\varphi'^2(\xi-\kappa)-\frac{3}{2}
\xi\frac{\varphi^2}{(1+x^2)^2}=-\frac{3}{(1+x^2)^2} +\frac{4
\epsilon  b^2}{ a_0^2 }\frac{3
x^2}{(1+x^2)^3}.\label{eqndimensional}\end{equation}
In general when $\xi\neq 0$ it is difficult to solve the above
equation. A way to (approximately) solve it is to assume the
parameter $\xi$ as a small perturbation, i. e., to apply a
perturbative analysis. It will be easier and more transparent to
split the analysis into three separated cases corresponding to
different field configurations (theories):

\vskip .2cm
I) A Gauss-Bonnet term plus a scalar field minimally
coupled to gravity ($\xi=0$) \cite{mg}.

\vskip .2cm II) A scalar field non-minimally coupled to gravity
without the Gauss-Bonnet term ($\epsilon=0$).

\vskip .2cm
III) The general case where $\epsilon \neq 0$ and the
small parameter $\xi\neq0$.

\vskip .2cm
\textbf{Case I) Minimally coupled
theory with Gauss-Bonnet term}

If one sets $\xi=0$ and considers the case where $a_0=2\sqrt{\epsilon }b$,
the background solution is \cite{mg1}:
\begin{equation}
\varphi(x)=\pm \varphi_0 \frac{x}{\sqrt{1+x^{2}}}+\varphi_{1}^{\pm},
\label{campo1}\end{equation}
\begin{equation}
V(\varphi)=V_0\left[ 3\left(\frac{\varphi -
\varphi_1^{\pm}}{\varphi_0}\right)^{4}-6\left(\frac{\varphi -
\varphi_1^{\pm}}{\varphi_0}\right)^2 +1\right],
\label{potencial1}\end{equation}
where $\varphi_0=\sqrt{\frac{3}{\kappa}}$ and  $V_0 =
\frac{3}{8\kappa\epsilon}$. The $\pm$ sign describes two possible
solutions of $\varphi$ with the constants $\varphi_{1}^{\pm}$. In
this case the 4D Planck mass does not depend explicitly on the
parameters of the scalar field, moreover, the definition
(\ref{masaplanck}) tell us that the addition of a Gauss-Bonnet term
to a model with a scalar field minimally coupled to gravity in the
action slightly reduces the value of the 4D Planck mass:
\begin{equation}
M_{\rm Pl}^2\sim \frac{4}{3}\frac{M^3 a_0^3} {b}\label{massp0}
\end{equation}
compared to its value when there is no Gauss-Bonnet term in the model
\begin{equation}
M_{\rm Pl}^2\sim \frac{2 M^3 a_0^3} {b}.
\label{massp}
\end{equation}
The above relation implies that the zero tensor mode is normalizable
and all the conditions (\ref{ligaduras}) are trivially satisfied.

As it can be seen from (\ref{potencial1}), the
self-interaction potential of the scalar field $V(\varphi)$
interpolates between two identical constant values.

In the cases II and III we will investigate solutions of
(\ref{eqndimensional}) when the non-minimal coupling is small
compared to the Gauss-Bonnet and Einstein-Hilbert contributions to
the action (\ref{action}), characterized by the parameters
$\epsilon$ and $\kappa,$ respectively. Thus, it is convenient to
make the quantities in (\ref{eqndimensional}) dimensionless since
one wishes to study the field configurations independently of the
choice of the units of measure. Let us consider the following
redefinitions: $\varphi=\frac{\phi}{\sqrt{\kappa}}$ and
$\varepsilon=\frac{\xi}{\kappa} \ll 1$. Under these new
dimensionless variables the field equation (\ref{eqndimensional})
takes the following form:
\begin{equation}
\!\!\!\!\!\!\!\!\!\!\!\!\!\!\!\!\!\!\!\!\!\!\!\!\varepsilon \phi \phi'' +\frac{2 \varepsilon x}{1+x^2} \phi
\phi'+\phi'^2 (\varepsilon - 1) - \frac{3}{2} \varepsilon
\frac{\phi^2}{(1+x^2)^2}=-\frac{3}{(1+x^2)^2} +\frac{4 \epsilon
b^2}{ a_0^2 } \frac{3
x^2}{(1+x^2)^3}.\label{ecu-adimensional}\end{equation}
In these two cases we will assume that the solution can be formally
expanded in powers of $\varepsilon$ \footnote{If the first-order
term $\varepsilon \phi_1(x)$ in (\ref{expregular}) is uniformly
``small'' with respect to the zeroth-order $\phi_0(x)$ when
$\varepsilon\rightarrow 0$ over some region in the domain of
variation (${\cal D}$) of $\phi$, in other words, if $\varepsilon
\phi_1(x)=o(\phi_0(x))$ when $\varepsilon \rightarrow 0$ over some
region $\in{\cal D}$, the expansion is called \emph{asymptotic
expansion} up to first order of the field on this region.}:
\begin{equation}
\phi=\phi_0(x) +\varepsilon \phi_1(x) + o(\varepsilon \phi_1(x)).
\label{expregular}\end{equation} By substituting this expansion into
(\ref{ecu-adimensional}) one can find approximate solutions of the
field equation. However, a subtlety arises when we try to apply this
naive procedure to (\ref{ecu-adimensional}) since in this case the
term containing the second derivative is multiplied by the small
perturbation parameter, namely, if one sets $\varepsilon=0$ in
(\ref{ecu-adimensional}) the differential equation that arises is no
longer of second order, but of first order. Therefore, it is not
possible to generate all the approximate solutions to the field
equation (\ref{ecu-adimensional}) with two \emph{arbitrary} boundary
conditions for $\phi$ at $x=\pm \infty$, because the differential
equations for $\phi_0$ and $\phi_1$ will be of first order. Thus,
the study of the boundary value problem poses a dilemma: there is
only one unknown arbitrary constant but there are two fixed boundary
conditions to be satisfied. This leads to what is generally known as
a \emph{singular perturbation} or \emph{boundary layer problem}
\cite{eckhaus,ferdinand,jean,murdock,holmes}.

In this kind of problems, in general, it is not possible to find a
single uniformly valid asymptotic expansion for the field, in
agreement with two arbitrary boundary conditions. What one can do is
to propose two different asymptotic expansions for regions that
contain the boundaries $x=-\infty$ and $x=\infty$, such that each
one of them is uniformly valid in the corresponding region.
Although, as we will see, in our case it is not enough to have two
asymptotic expansions: we shall require another asymptotic expansion
in the neighborhood of the origin ($x=0$).

In addition to the mentioned complications, in this kind of problems
usually there is a region of rapid variation of the field and/or of
its first derivatives, known as the boundary layer. In this region
the term that contains the second derivative in the field equation
is no longer negligible, thus the expansion (\ref{expregular}) and
the corresponding solutions are  not valid in this region.

In order to find an asymptotic expansion uniformly valid in a
boundary layer located at $x_b$, characterized in size by the
function $\delta(\varepsilon)$, it is suitable to magnify this
region by rescaling the variable $x$ with the aid of the stretched
or boundary layer variable:
\begin{equation}
\zeta=\frac{x-x_b}{\delta(\varepsilon)},\quad
\mbox{with}\quad\delta(\varepsilon)=o(1)\quad\mbox{when}\quad\varepsilon\rightarrow
0.\end{equation} In terms of this variable the field transforms
into:
\begin{equation}
\phi(x)=\phi(x_b +\delta(\varepsilon) \zeta)\equiv\Phi(\zeta).
\label{campo_layer}\end{equation}
The next step is to consider a different expansion for $\Phi(\zeta)$
that we hope to be uniformly valid on the boundary layer region. How
to choose $\delta(\varepsilon)$ in order to satisfy this condition
will be discussed later.

On the other hand, in order to obtain an approximation on the whole
domain ${\cal D}$, it is necessary to know how to match two
different adjacent asymptotic expansions. The principal
\emph{hypothesis} of this method is to assume that there is an
intermediate region where two different asymptotic expansions give
the same result. By following \cite{ferdinand}, the idea consists in
defining on the intermediate domain a new stretched variable:
\begin{eqnarray}\label{variableinter}
&&\zeta_0=\frac{x-x_b}{\delta_0(\varepsilon)},\,\,\,\mbox{with}\,\,\,
\delta_0(\varepsilon)=o(1),\,\,\,\delta(\varepsilon)=o(\delta_0(\varepsilon))\,\,\,
\mbox{when}\,\,\,\varepsilon\rightarrow 0,\end{eqnarray} and
rewriting the asymptotic expansions in terms of $\zeta_0$ up to some
order\footnote{Not necessarily the same order for both asymptotic
expansions.} and then, after expanding both of them, asking for
their equality in terms of an arbitrary function
$\delta_0(\varepsilon)$ under the restrictions written in
(\ref{variableinter}). This matching procedure serves us to fix the
constants that appear in the expansion for $\Phi(\zeta)$ on regions
which are not connected with the boundary conditions.

A description of the solution on certain regions consists of two
expansions which must be combined to form a {\it composite
expansion}. This is done by adding the expansions and then
subtracting the part that is common to both, yielding an
approximation to the solution on the above mentioned region
\cite{holmes}. After determining all the composite expansions
uniformly valid on different regions, the resulting approximation to
the solution valid on the whole interval can be obtained by joining
together all of them.

\vskip .2cm
\textbf{Case II) Non-minimally coupled
theory without Gauss-Bonnet term}

In this case the equation (\ref{ecu-adimensional}) can be written as
\begin{equation}
\varepsilon\phi\phi''+\frac{2\varepsilon x}{1+x^2}\phi\phi' +\phi'^2 (\varepsilon-1)-
\frac{3}{2}\varepsilon\frac{\phi^2}{(1+x^2)^2}=-\frac{3}{(1+x^2)^2},\label{eqnnonminimal}
\end{equation}
and the consistency conditions (\ref{ligaduras}) can be reduced to
the following pair of constraints: $L(\phi)=1-\frac{\varepsilon}{2}
\phi^2>0$ and  $M_{\rm Pl}^{2} > 0$.

\vskip .2cm
\textbf{IIa) Case without boundary
conditions}

\begin{figure}[htb]
\begin{center}
\includegraphics[width=8cm]{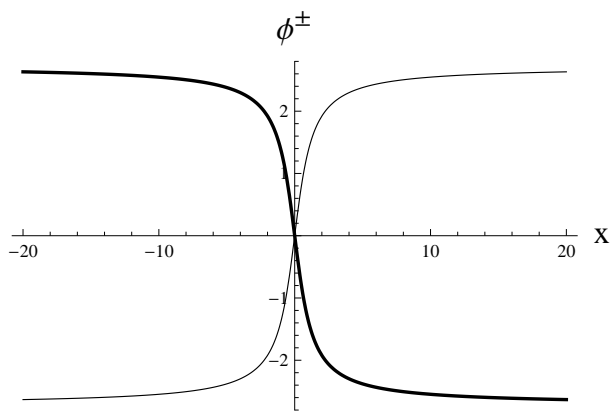}
\end{center}
\caption{Graph of the function $\phi(x)$ up to first order in
$\varepsilon$ for case IIa). In the figure we set $\varepsilon=0.01$
and $A^{\pm}_0=A^{\pm}_1=0$. The thin line represents
$\phi^{+}=\phi^{+}_0+\varepsilon \phi^{+}_1$ and the thick one
describes $\phi^{-}=\phi^{-}_0+\varepsilon \phi^{-}_1$.}
\label{campoescalarnominimo}
\end{figure}

\begin{figure}[htb]
\begin{center}
\includegraphics[width=8cm]{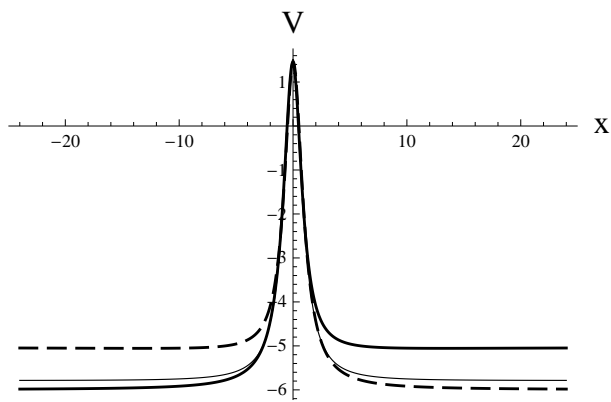}
\end{center}
\caption{Self-interaction potential $V(\phi(x))$ up to first order
in $\varepsilon$ for case IIa) (in all cases we have chosen $\kappa=1$, $a_0=1$, $b=1$ and
$\varepsilon=0.01$). The following values of the constant parameters
have been chosen: for $\phi^{+}_0$, $A^{\pm}_0=0$ -- thin line,
$A^{+}_0=3$ -- thick line, while the dashed line corresponds to the
choice $A^{-}_0=3$, for $\phi^{-}_0$.}
\label{figVnominimo}\end{figure}

\begin{figure}[htb]
\begin{center}
\includegraphics[width=8cm]{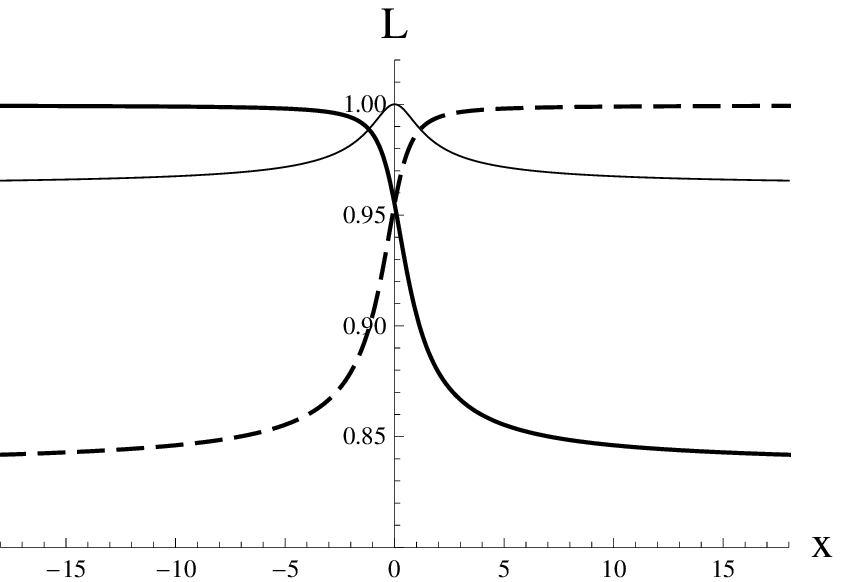}
\end{center}
\caption{Behaviour of the function $L(\phi)$ -- up to first order in
$\varepsilon$ -- vs the variable $x$ for the case IIa). We have
chosen three different sets of values of the free constants
($\varepsilon=0.01$ in all cases): i) $A^{\pm}_0=0$, $\phi^{\pm}_0$
-- thin line, ii) $A^{+}_0=3$, $\phi^{+}_0$ (thick line), and iii)
$A^{-}_0=3$, $\phi^{-}_0$ -- dashed line.}
\label{figLnominimo}\end{figure}

These last restrictions do not require to fix any initial or
boundary conditions. Thus, to begin with, let us study the class of
approximate solutions generated by (\ref{expregular}). By
substituting this expansion into (\ref{eqnnonminimal}) it is not
difficult to obtain the equations for the first and second
approximations, $\phi_0(x)$ and $\phi_1(x),$ respectively:
\begin{eqnarray}
&&\phi_0'^2-\frac{3}{(1+x^{2})^{2}}=0, \label{eqnceronominimo}\\
&&2\phi_0'\phi_1'-\frac{2x}{1+x^{2}}\phi_0\phi_0'-\phi_0\phi_0''-\phi_0'^2+
\frac{3}{2}\frac{\phi_{0}^{2}}{(1+x^{2})^{2}}=0.\label{eqnunonominimo}\end{eqnarray}
The above equations give us the following solutions
\begin{eqnarray}
\phi^{\pm}_0&=&\pm \sqrt{3}\arctan(x) +A^{\pm}_0,\label{ceronominimo}\\
\phi^{\pm}_1&=&\pm \frac{\sqrt{3}}{2}\arctan(x) \mp\frac{\sqrt{3}}{4}(A^{\pm}_{0})^{2}
\arctan(x)\mp \frac{\sqrt{3}}{4} \arctan^{3}(x) \nonumber \\
&&-\frac{3}{4}A^{\pm}_0\arctan^{2}(x)+A^{\pm}_1,
\label{unonominimo}\end{eqnarray} where the signs $\pm$ mean two
possible solutions for each order of approximation, and $A^{\pm}_0$
and $A^{\pm}_1$ are arbitrary constants. One can check that the
above solutions generate an asymptotic expansion
$\phi^{\pm}=\phi^{\pm}_0+\varepsilon \phi^{\pm}_1$ on the whole
domain ${\cal D}=(-\infty,\infty) $ of the field variable.
In Fig. \ref{campoescalarnominimo} it is shown that
there are configurations of the scalar field with a kink-like
behaviour. Moreover, in contrast to the previous case, the
self-interaction potential, in general, interpolates between two
different negative constant values as shows the Fig.
\ref{figVnominimo}; these asymptotic values ​​can be written as
follows:
\begin{eqnarray}
V^{+}(\pm \infty) \sim \frac{3b^2}{4 a_{0}^{2} \kappa}\left[ -8 + \varepsilon\left(2A_0^{+} \pm \sqrt{3}\pi\right)^2\right],\nonumber \\
V^{-}(\pm \infty) \sim \frac{3b^2}{4 a_{0}^{2}
\kappa}\left[-8+\varepsilon\left(2A_0^{-} \mp
\sqrt{3}\pi\right)^2\right],\nonumber
\end{eqnarray}
where $V^{+}$ and  $V^{-}$ correspond to the solutions
$\phi^{+}$ and $\phi^{-}$ respectively. This asymmetric asymptotic behaviour of the matter
energy density of the system may seem surprising if one takes
into account the fact that the space-time is asymptotically
$AdS_5$. However, the uneven
character of the energy distribution of the scalar matter at
$-\infty$ and $+\infty$ is compensated by its non-minimal coupling
to gravity, rendering an asymptotically $AdS_5$ space-time which is
even with respect to the fifth coordinate. Furthermore, up to first
order in $\varepsilon$ the 4D Planck mass is
\begin{equation}
M_{\rm Pl}^{2} \sim\frac{a_{0}^{3} M^{3}}{b}\left\lbrace 2-
\varepsilon\left[\frac{3}{4}\pi^2+\left(A_{0}^{\pm}\right)^2-6\right]
\right\rbrace. \nonumber
\end{equation}
As we expected from (\ref{masaplanck}), the above formula tells us
that the effect of the non-minimal coupling between the scalar field
and gravity reduces the value of the 4D Planck mass compared to its
value when there is only minimal coupling. By taking into account
the requirements $L(\varphi)>0$ and  $M_{\rm Pl}^{2} > 0$ mentioned
above, one can show that the values of $\varepsilon$ are bounded
\begin{equation}\nonumber
\varepsilon <
\frac{2}{\left(\sqrt{3}\frac{\pi}{2}+|A_{0}^{\pm}|\right)^2}
\end{equation}
In Fig. \ref{figLnominimo} it is shown the function
$L(\phi)=1-\frac{\varepsilon}{2} \phi^2$ for several sets of values
of the parameters. As one can see, the condition $L(\phi)>0$ is
obviously satisfied.

As one can see from (\ref{ceronominimo}) and (\ref{unonominimo}),
after fixing the sign of the solutions, only one free boundary
condition is left. In consequence, it is not possible to generate
all of the approximate solutions of (\ref{eqnnonminimal}) by using
only one expansion of $\phi$.

For arbitrary boundary conditions it is difficult to solve
(\ref{eqnnonminimal}). Hence, here we solve the field equation just
for the case where the field
vanishes at $x=\pm \infty$. These boundary conditions are
interesting because these are the only boundary conditions that will
appear in the next relevant case where $\xi\neq 0$ and $\epsilon
\neq 0$.

\vskip.2cm
\textbf{IIb) Case with imposed boundary
conditions}

Let us consider the equation (\ref{eqnnonminimal}) under the
following boundary conditions:
\begin{equation}
\phi(-\infty)=\phi(\infty)=0. \label{frontnominimo}\end{equation}
First of all, it is easy to show that if $\phi(x)$ is a solution of
(\ref{eqnnonminimal}) under (\ref{frontnominimo}), then, $\phi(-x)$
is a solution of the field equation too, with the same boundary
conditions. This implies that the solution of the field equations
under the conditions (\ref{frontnominimo}) is an even function.

Let us solve perturbatively (\ref{eqnnonminimal}) under the boundary
conditions (\ref{frontnominimo}). The expansion outside the boundary
layers is described by the solutions (\ref{ceronominimo}) and
(\ref{unonominimo}). As one can see, it is not possible to construct
a single expansion which satisfies both boundary conditions at the
same time. Therefore, we need to find asymptotic expansions valid on
the boundaries. In general, this is a difficult task, but in our
case we shall see that if one chooses any expansion $\phi^{+}$ or
$\phi^{-}$ for a domain that contains the boundary $x=+\infty$ and
the remaining solution for a region that contains the point
$x=-\infty$, the constants that appear in the solutions can be fixed
in such a way that the boundary conditions can be completely
satisfied.

In what follows, we will denote with the index $\alpha$ the case
where $\phi^{-}$ is defined on the region that contains the boundary
$x=-\infty$ and $\phi^{+}$ on the region that contains the boundary
$x=+\infty$, otherwise the subscript $\beta$ will be used. To begin
with, let us discuss in detail the case $\alpha$. The other case is
similar, therefore, only the final results will be presented.

By setting $A^{+}_0=A^{-}_0=A_0=-\frac{\sqrt{3}}{2} \pi$ and
$A^{+}_1=A^{-}_1=A_1=\frac{\sqrt{3}}{4}
\pi\left(\frac{\pi^{2}}{8}-1\right)$, we can write the approximate
solution to (\ref{eqnnonminimal}) under (\ref{frontnominimo}) for
regions that contain the boundaries in the form:
\begin{eqnarray}
\!\!\!\!\!\!\!\!\!\!\!\!\!\phi^{-}_\alpha&=&-\sqrt{3}\arctan(x) +A_0 -\frac{\sqrt{3}}{4}
\varepsilon[2\arctan(x) -A_{0}^{2}\arctan(x)-\arctan^{3}(x) \nonumber \\
\!\!\!\!\!\!\!\!\!\!\!\!\!&& +\sqrt{3} A_0\arctan^{2}(x)]+\varepsilon A_1, \quad
\mbox{for all}\,x
\in {\cal D}^{-},\label{cornermenos}\\
\!\!\!\!\!\!\!\!\!\!\!\!\!\phi^{+}_\alpha&=&\sqrt{3}\arctan(x) +A_0 +\frac{\sqrt{3}}{4}
\varepsilon[2\arctan(x) -A_{0}^{2}\arctan(x)-\arctan^{3}(x)  \nonumber \\
\!\!\!\!\!\!\!\!\!\!\!\!\!&& -\sqrt{3} A_0\arctan^{2}(x)]+\varepsilon A_1, \quad
\mbox{for all}\,x \in {\cal D}^{+}, \label{cornermas}
\end{eqnarray}
where the domains ${\cal D}^{-}$ and ${\cal D}^{+}$ contain the
points $x=-\infty$ and $x=+\infty,$ respectively.

If one naively supposes that there is no intermediate boundary
layers between the two boundary regions, the next step is to find a
common region where the two expansions are valid. In our case, it is
not difficult to find it because $\phi^{-}_{\alpha}$ and
$\phi^{+}_{\alpha}$ can be matched order by order at $x=0$. Thus,
the domains ${\cal D}^{\pm}$ can be taken as  ${\cal
D}^{-}=(-\infty,0]$ and ${\cal D}^{+}=[0,\infty),$ respectively.
These two domains allow us to define a general solution (valid on
the entire domain) for the field. As one can see from
(\ref{cornermenos}) and (\ref{cornermas}), the latter solution is
continuous but its derivative is not continuous at $x=0$. In other
words, on a neighborhood of the origin the first derivative of the
field changes ``rapidly" from negative to positive values. Therefore
we have a boundary layer behaviour of the first derivative of the
field on a neighborhood of the point $x=0$.

The above analysis tell us that our initial hypothesis about the
absence of intermediate boundary layers is false, then, we need to
find an asymptotic expansion valid on a neighborhood of the origin.
As we mentioned before, it is convenient to define a new variable
\begin{equation}
\zeta=\frac{x}{\delta(\varepsilon)},\quad
\mbox{with}\quad\delta(\varepsilon)=o(1), \quad \mbox{when}
\quad\varepsilon\rightarrow 0,\label{layervariable}\end{equation}
and to perform a new expansion for $\phi_{\alpha}(x)=\Phi(\zeta)$
over the domain that contains the origin $x=0$:
\begin{equation}
\Phi(\zeta)=\Phi_0(\zeta) +\varepsilon \Phi_1(\zeta).
\label{expsingular}\end{equation} In terms of the new variable
(\ref{layervariable}) the equation (\ref{eqnnonminimal}) can be
rewritten as:
\begin{eqnarray}
\!\!\!\!\!\!\!\!\!\!\!\!\!&&\varepsilon \Phi \frac{d^2 \Phi}{d \zeta^2}+2 \varepsilon
\delta^2 \frac{\zeta}{1+(\delta \zeta)^2} \Phi \frac{d \Phi}{d \zeta}
+(\varepsilon - 1)\left(\frac{d\Phi}{d \zeta}\right)^{2} -\frac{3}{2}
\varepsilon \delta^2 \frac{\Phi^2}{(1+(\delta \zeta)^2)^2}\\ \nonumber
\!\!\!\!\!\!\!\!\!\!\!\!\!&&= -\frac{3\delta^2}{(1+(\delta \zeta)^2)^2}.\label{ecu-nmlayer}
\end{eqnarray}
 By substituting (\ref{expsingular}) into the previous equation, we get
\begin{eqnarray}
\!\!\!\!\!\!\!\!\!\!\!\!\!\!\!\!\!\!\!\!\!\!\!\!\!\!&&-\left(\frac{d\Phi_0}{d \zeta}\right)^2+\varepsilon\left[\Phi_0
\frac{d^2\Phi_0}{d \zeta^2}-2 \frac{d\Phi_0}{d
\zeta}\frac{d\Phi_1}{d \zeta}+ \left(\frac{d\Phi_0}{d
\zeta}\right)^2\right]+\varepsilon \delta^2\left[2 \zeta \Phi_0
\frac{d\Phi_0}{d \zeta}-\frac{3}{2}\left(\Phi_0\right)^2\right]+
\nonumber \\
\!\!\!\!\!\!\!\!\!\!\!\!\!\!\!\!\!\!\!\!\!\!\!\!\!\!&&\varepsilon^2\left[\Phi_0 \frac{d^2\Phi_1}{d
\zeta^2}+2\frac{d\Phi_0}{d\zeta}\frac{d\Phi_1}{d \zeta}+\Phi_1
\frac{d^2\Phi_0}{d \zeta^2}- \left(\frac{d\Phi_1}{d
\zeta}\right)^2\right]+\cdots=-3\delta^2+\cdots.\label{eqn-desarrollolayer}\end{eqnarray}
In order to determine $\Phi_0$ and $\Phi_1$ we need to know how to
choose $\delta(\varepsilon)$. The idea is to make a well-balanced
choice of the variable $\zeta$ or, equivalently, of the
$\delta(\varepsilon)$ function, such that the equations for the
approximations $\Phi_0$ and $\Phi_1$ contain as much information as
possible as $\varepsilon \rightarrow 0$. This selection is called
the \emph{distinguished limit} of the expansion (\ref{expsingular})
\cite{eckhaus}--\cite{murdock}. Let us put it in different words: if
we choose $\delta(\varepsilon)$ in a way different from the
distinguished limit, all the terms that arise in the equations for
$\Phi_0$ and $\Phi_1$ will be contained in the equations for these
functions in the distinguished limit.

In our case we can show that $\delta=\varepsilon$ corresponds to the
distinguished limit of (\ref{expsingular}), therefore, by using
(\ref{eqn-desarrollolayer}) the equations for the zero and first
order approximations of $\Phi$ on the boundary layer region are:
\begin{eqnarray}
\frac{d\Phi_0}{d \zeta}&=&0, \label{ecu0layer}\\
\Phi_0 \frac{d^2 \Phi_1}{d \zeta^2}-\left(\frac{d\Phi_1}{d
\zeta}\right)^2&=&-3. \label{ecu1layer}\end{eqnarray} By solving the
above system of equations, the field $\Phi(\zeta)$ on the boundary
layer domain adopts the form:
\begin{equation}
\Phi(\zeta)=\Phi_0+\varepsilon\left[c_2-\Phi_0\ln\cosh\left(
\frac{\sqrt{3}}{\Phi_0}(\zeta+c_1)\right)\right],\label{sollayer}\end{equation}
where $c_1$, $c_2$ and $\Phi_0$ are constants. These constants can
be determined by matching the expansion of (\ref{cornermenos}), up
to some order, with the above solution on a negative neighborhood of
the point $x=0,$ and by matching the expansion of the solution
(\ref{cornermas}), up to some order, with (\ref{sollayer}) on a
positive neighborhood of the point $x=0$, respectively. Let us next
define the intermediate stretched variable:
\begin{eqnarray}\label{vintermedia}
&&\zeta_0=\frac{x}{\delta_0(\varepsilon)},\,\,\,\mbox{with}
\,\,\,\delta_0(\varepsilon)=o(1),\,\,\,\varepsilon=o(\delta_0(\varepsilon))
\,\,\,\mbox{when}\,\,\,\varepsilon\rightarrow 0.\end{eqnarray} By
expanding the zeroth order approximations of (\ref{cornermenos}) and
(\ref{cornermas}) and the boundary layer solution (\ref{sollayer})
in terms of $\delta_0(\varepsilon)$, we get\footnote{In order to
include the first order approximation $\phi_{1\,\alpha}^{\pm}(x)$ in
the matching process, it is necessary to calculate the contribution
of $\varepsilon^2 \Phi_2(\zeta)$ to the expansion
(\ref{expsingular}). However, this case is analytically more complex
and, thus, will not be considered here.}:
\begin{eqnarray}
&&\phi^{\pm}_{\alpha}\sim \pm \sqrt{3}\delta_0 \zeta_0-\sqrt{3}\frac{\pi}{2},\nonumber\\
&&\Phi^{\pm}\sim \Phi_0 \pm \sqrt{3}\delta_0 \zeta_0+\varepsilon(\pm
\sqrt{3}c_1+c_2+ \Phi_0 \ln 2),\label{aproxlayer}\end{eqnarray}
where in the $\pm$ symbol, the $-$ sign stands for quantities
defined on a negative neighborhood of the origin, similarly, the $+$
sign denotes quantities defined on a positive neighborhood of the
origin.

By matching the above expansions, the following values for the
constants $\Phi_0$, $c_1$ and $c_2 $ are obtained
\begin{eqnarray}
&&\Phi_0=-\sqrt{3}\frac{\pi}{2},\nonumber\\
&&c_1=0,\nonumber\\
&&c_2=\sqrt{3}\frac{\pi}{2}\ln 2.\nonumber\end{eqnarray}
Notice that, up to this point, the description of our approximate
solution is realized by three well-defined asymptotic expansions
(which form two pairs), that can be joined together to get a
solution that is valid on the entire domain ${\cal
D}=\{-\infty<x<+\infty\}$. One pair of these asymptotic expansions
is uniformly valid on the region $\{x\le 0\}\in {\cal D}$, while the
other pair is uniformly valid on the region $\{x\ge 0\}\in {\cal
D}$. Thus, we can construct two {\it composite expansions} defined
on the regions $x\le 0$ and $x\ge 0$, respectively, that will be
collected to generate an approximate solution defined on the whole
domain of variation ${\cal D}$.

As we mentioned above, the composite expansion for the region $x\le
0$ (or $x\ge 0$) is obtained by adding two asymptotic expansions
uniformly valid on the corresponding region and expressed in terms
of the same variable, and then by subtracting the part that is
common to both of them. In our case it is convenient to construct
the composite expansions for the positive and negative regions
($x\le 0$ and $x\ge 0$) separately.

In other words, the composite expansion (either for $x\le 0$ or
$x\ge 0$) can be expressed as follows
\begin{equation}
\phi_{c}^{\alpha}(x)=\phi_{0}^{\alpha}(x)+\Phi
\left(\frac{x}{\varepsilon}\right)-\phi_{com}(x),\nonumber
\end{equation} where $\phi_{com}(x)$ is the common part to both
asymptotic expansions. It should be noticed that in our case, by
virtue of (\ref{aproxlayer}), the common functions for the
expansions on the $x\le 0$ and $x\ge 0$ regions can be described in
a single functional form
\begin{equation}
\phi_{com}(x)=\sqrt{3}\left(|x|-\frac{\pi}{2}\right).\end{equation}
\begin{figure*}[htb]
\begin{center}
\includegraphics[width=16cm]{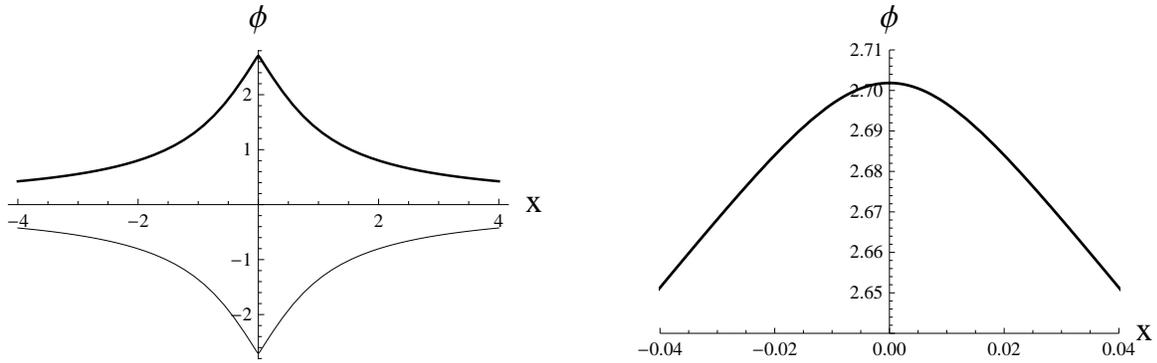}
\end{center}
\caption{The graph (a) shows the solutions for the field $\phi$ in
the case IIb); the thin line represents $\phi^{\alpha}(x)$ and the
thick line represents $\phi^{\beta}(x)$. Both  profiles
asymptotically tend to zero. In graph (b) we represent the profile
of $\phi^{\beta}(x)$ in the neighborhood of the point $x=0$. In
these figures we set $\varepsilon=0.01$.}
\label{fcompnomin}\end{figure*}

\begin{figure}[htb]
\begin{center}
\includegraphics[width=16cm]{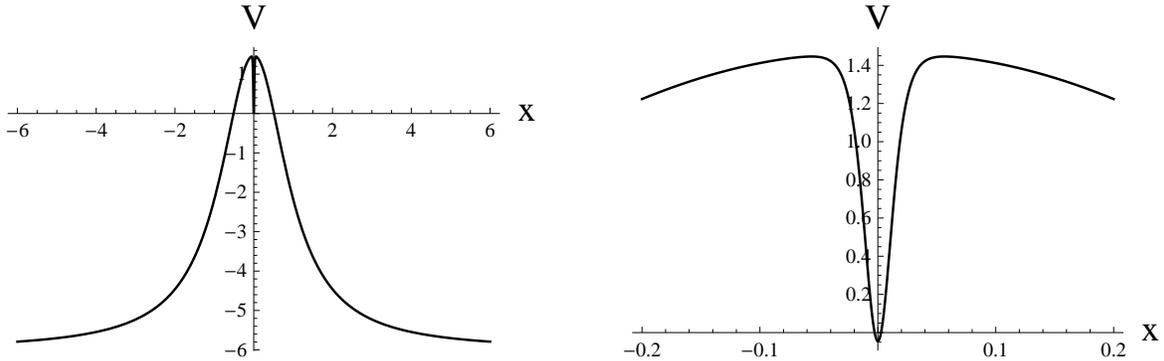}
\end{center}
\caption{The graph (a) shows the self-interaction potential vs $x$
for the case IIb). In graph (b) we  zoomed the self-interaction
potential in the neighborhood of the point $x=0$. In these figures
we set $\kappa=1$, $a_0=1$, $b=1$ and $\varepsilon=0.01$.} \label{potencialcomnomin}
\end{figure}

\begin{figure*}[htb]
\begin{center}
\includegraphics[width=16cm]{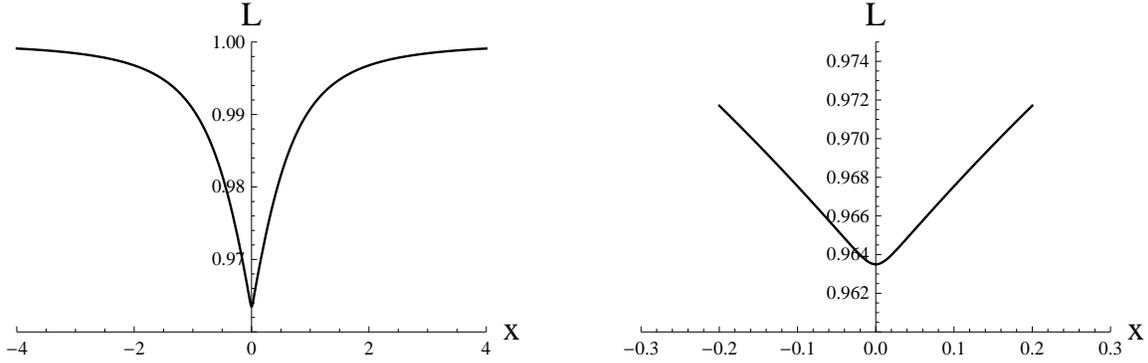}
\end{center}
\caption{ The graph (a) shows the function $L(\phi)$ vs $x$ for the
case IIb). In graph (b) we zoomed  this function in the neighborhood
of the point $x=0$. In these figures we set
$\varepsilon=0.01$.}\label{Lfuncionlayer1}\end{figure*}

Thus, the solution is:
\begin{equation}
\!\!\!\!\!\!\!\!\!\!\!\!\!\!\!\phi^{\alpha}(x)=\phi(x)=\sqrt{3}\left\lbrace\arctan|x|-
\left(|x|+\frac{\pi}{2}\right)+\frac{\pi}{2}\varepsilon\ln\left[2\mbox{cosh}
\left(\frac{2x}{\pi
\varepsilon}\right)\right]\right\rbrace.\end{equation} By following
the same scheme described above, we can show that
$\phi^{\alpha}(x)=-\phi^{\beta}(x)$.

As it is shown in Fig. \ref{fcompnomin}a, both solutions are
bounded, continuous even functions, with continuous derivatives on
the whole domain ${\cal D}$. In Fig. \ref{fcompnomin}b we have
zoomed $\phi^{\beta}(x)$ on a region near to the origin of
coordinates to appreciate more clearly the smooth behaviour of this
profile. Furthermore, like in the case I, the
self-interaction potential interpolates between two
identical negative constant values given by (see
Fig. \ref{potencialcomnomin}a):
$$
V(\infty) \sim -\frac{6 b^2}{a_{0}^{2} \kappa}.
$$
On the other hand, the 4D Planck mass is
\begin{equation}
M_{\rm Pl}^{2} \sim\frac{a_{0}^{3} M^{3}}{b}\left\lbrace 2-
3\varepsilon\left(\pi-2\right) \right\rbrace,
\label{massp2}
\end{equation}
again, the non-minimal coupling effects diminishes $M_{\rm Pl}^{2}$
with respect to a simpler model where the field is minimally coupled
to gravity and the Gauss-Bonnet term is absent. By using
$L(\varphi)>0$ and the above obtained result (\ref{massp2}), the
restriction for the values of $\varepsilon$ is
$$\varepsilon<\frac{8}{3 \pi^2}.$$

In Fig. \ref{Lfuncionlayer1}a it is shown the function $L(\phi)$ for
both solutions. As we wished, the condition $L(\phi)>0$ is
satisfied.

\vskip .2cm
\textbf{Case III) Non-minimal coupling
and Gauss-Bonnet term}

Like in the first case, the relation $a_0=2\sqrt{\epsilon }b$ is
imposed, then, the field equation (\ref{ecu-adimensional}) is
transformed into:
\begin{equation}
\varepsilon\phi\phi''+\frac{2\varepsilon x}{1+x^2}\phi\phi'
+\phi'^2(\varepsilon-1)-\frac{3}{2}\varepsilon\frac{\phi^2}{(1+x^2)^2}=
-\frac{3}{(1+x^2)^3},\label{eqngaussnm}\end{equation}
In this case we shall modify the condition $\langle \psi|\psi\rangle
> 0$ of (\ref{ligaduras}) by a stronger and more tractable one,
where the weight function $r/s$ is not negative. Under this
redefinition of (\ref{ligaduras}) the admissible profiles for $\phi$
are constrained by the following condition:
\begin{equation}
f(x)=\frac{1}{1+x^2} -\frac{\varepsilon}{2} \phi^2 \geq 0,
\label{condgaussnm}\end{equation} then, when $x \rightarrow \infty$
the field vanishes. Based on this fact, we will study approximate
solutions to (\ref{eqngaussnm}) under the boundary conditions:
\begin{equation}
\phi(-\infty)=\phi(\infty)=0.\label{condfrontgaussnm}\end{equation}
Of course, this boundary value problem is weaker than
(\ref{eqngaussnm}) under the restrictions (\ref{ligaduras}) and
(\ref{condgaussnm}). Thus, after finding the solutions, it is
necessary to check whether such restrictions are fulfilled.

As mentioned above, we will find solutions of (\ref{eqngaussnm}) in
terms of the expansion (\ref{expregular}). By substituting such an
expansion into (\ref{eqngaussnm}), the equations for the zeroth and
first approximations become:
\begin{eqnarray}
&&\phi'^2_0-\frac{3}{(1+x^2)^3}=0, \nonumber\\
&&2\phi'_0 \phi'_1 - \phi_{0}'^2 -\phi_0 \phi''_0 - \frac{2
x\phi_0\phi'_0}{1+x^2} + \frac{3}{2}
\frac{\phi_0^2}{(1+x^2)^2}=0,\nonumber\end{eqnarray} and possess the
following solutions:
\begin{eqnarray}
\!\!\!\!\!\!\!\!\!\!\!\!\!\phi_{0}^{\pm}&=&\pm \sqrt{3}\frac{x}{\sqrt{1+x^2}} + B_{0}^{\pm}, \nonumber\\
\!\!\!\!\!\!\!\!\!\!\!\!\!\phi_{1}^{\pm} &=&\frac{1}{4\sqrt{3}} \left\lbrace
\frac{\pm21x}{\sqrt{1+x^2}}
\mp3\left[5+(B_{0}^{\pm})^2\right]\mbox{arcsinh({\it
x})}-4\sqrt{3}B_{0}^{\pm}
\ln(1+x^2)\right\rbrace+B_{1}^{\pm},\nonumber\end{eqnarray} where
$B_{0}^{\pm}$ and $B_{1}^{\pm}$ are constants. As in the previous
case, for each order of perturbation we have two solutions
characterized by the signs $\pm$. Since we have well-defined
boundary conditions, we must set the values of the above constants.
Again, with a single solution it is not possible to satisfy both
conditions simultaneously. Then, motivated by the case II we have
two possibilities of finding asymptotic expansions which are valid
on the boundaries. A first one where $\phi^{-}$ is valid on a region
that contains $x=-\infty$, and $\phi^{+}$ is valid on a region that
contains $x=+\infty$. By following the same notation used in the
case II, we will denote this choice by an $\alpha$ index. The second
one corresponds to choosing the expansions in reverse order and will
be denoted by a $\beta$ index. Let us consider the first case in
detail.

Like in the case II, a boundary layer behaviour of the first
derivative on the neighborhood of the origin arises. In order to
overcome this problem we repeat the same procedure applied to the
case II: we perform the change of variable (\ref{layervariable}) and
express (\ref{eqngaussnm}) in terms of it:
\begin{eqnarray}
\!\!\!\!\!\!\!\!\!\!\!\!\!\!&&\varepsilon \Theta \frac{d^2 \Theta}{d \zeta^2}+2
\varepsilon\delta^2\frac{\zeta}{1+(\delta \zeta)^2} \Theta \frac{d
\Theta}{d\zeta} +(\varepsilon - 1)\left(\frac{d\Theta}{d
\zeta}\right)^{2} -\frac{3}{2} \varepsilon \delta^2
\frac{\Theta^2}{(1+(\delta\zeta)^2)^2}\\ \nonumber
\!\!\!\!\!\!\!\!\!\!\!\!\!\!&&=-\frac{3 \delta^2}{(1+(\delta
\zeta)^2)^3},\label{ecu-gbnmlayer}\end{eqnarray}
where
$\Theta(\zeta)$ is the field written in terms of the $\zeta$
variable. When one compares (\ref{ecu-nmlayer}) to the above
equation, the only difference appears in the right hand side of the
equations, but, after expanding (\ref{expregular}) on the boundary
layer region, we only consider the first-order term in $\delta$.
Thus, the equations for the two first approximations of the field on
the boundary layer domain are the same that we obtained in the
previous case and the solution takes the form:
\begin{equation}
\Theta(\zeta)=\Theta_0+\varepsilon\left\{d_2-\Theta_0\ln\cosh\left[
\frac{\sqrt{3}}{\Phi_0}(\zeta+d_1)\right]\right\},\label{sollayer1}\end{equation}
where $\Theta_0$ is constant and describes the zeroth order
approximation of the field and $d_1$, $d_2$ are arbitrary constants.
By matching the three expansions we have:
\begin{eqnarray}
&&\Theta_0=-\sqrt{3},\nonumber\\
&&d_1=0,\nonumber\\
&&d_2=\sqrt{3}\ln 2.\nonumber\end{eqnarray}
\begin{figure}[htb]
\begin{center}
\includegraphics[width=8cm]{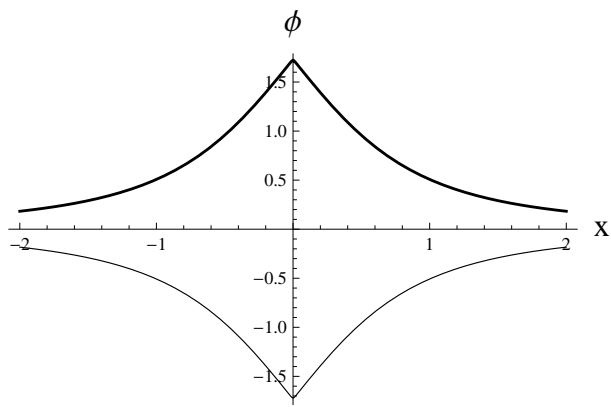}
\end{center}
\caption{The solutions for the field $\phi$ in the case III; the
thin line represents $\phi^{\alpha}(x)$ and the thick line denotes
$\phi^{\beta}(x)$. Both profiles tend asymptotically to zero. In the
figure we chose $\varepsilon=0.01$.} \label{campogbnm}\end{figure}

\begin{figure}[htb]
\begin{center}
\includegraphics[width=8cm]{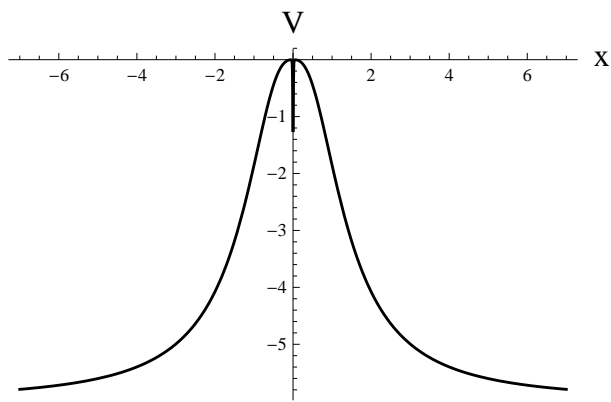}
\end{center}
\caption{Self-interaction potential vs $x$ for the case III
($\kappa=1$, $a_0=1$, $b=1$ and $\varepsilon=0.01$).}
\label{potencialcomgbnm}
\end{figure}
\begin{figure}[htb]
\begin{center}
\includegraphics[width=8cm]{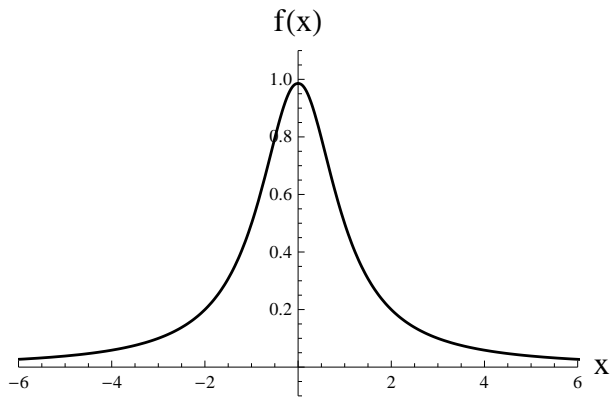}
\end{center}
\caption{The figure shows the fulfillment of the consistency
condition (\ref{condgaussnm}). As before we choose
$\varepsilon=0.01$.}\label{consistenciagbnm}\end{figure}

Finally, the solution for the field is
\begin{equation}\label{global1}
\phi^{\alpha}(x)=\phi(x)=\sqrt{3}\left\lbrace\frac{|x|}{\sqrt{1+x^2}}
- 1 -|x|+\varepsilon\ln\left[
2\mbox{cosh}\left(\frac{x}{\varepsilon}\right)\right]\right\rbrace.\end{equation}
By applying the same procedure as in case IIb), one can show that
$\phi^{\beta}(x)=-\phi^{\alpha}(x)$. From Fig. \ref{campogbnm} we
see that both profiles of the field are continuous, bounded even
functions, with continuous derivatives on the whole domain.
The asymptotic value of the self-interaction
potential reads:
$$
V(\infty) \sim -\frac{6 b^2}{a_{0}^{2} \kappa},
$$
and its profile is shown in Fig. \ref{potencialcomgbnm}, it is
similar to the one illustrated in Fig. \ref{potencialcomnomin}a.
Furthermore, the value of $M_{\rm Pl}^{2}$ is
\begin{equation}
M_{\rm Pl}^{2} \sim\frac{a_{0}^{3} M^{3}}{b}\left(\frac{4}{3}-
\varepsilon\right).
\nonumber
\end{equation}
By comparing the previous expression to (\ref{massp0}) one can see
the effect of the non-minimal coupling: it reduces the value of the
4D Planck mass compared to its value in a model in which the scalar
field is minimally coupled to gravity and there is a Gauss-Bonnet
term on the bulk. The consistency conditions (\ref{ligaduras})
together with the modification (\ref{condgaussnm}) imply that
\begin{equation}\nonumber
\varepsilon < \frac{2}{3}.
\end{equation}
As one can see from Fig. \ref{consistenciagbnm}, the consistency
condition (\ref{condgaussnm}) is fully satisfied for these constrained values
of $\varepsilon$.

\begin{figure}[htb]
\begin{center}
\includegraphics[width=8cm]{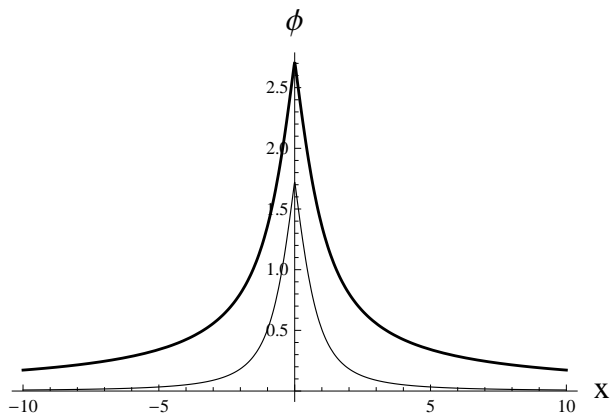}
\end{center}
\caption{Comparison of the solutions. The thin line represents
$\phi^{\beta}(x)$ for the case IIb) and the thick line represents
$\phi^{\beta}(x)$ for the case III. In both profiles we have
arbitrarily set $\varepsilon=0.01$.}
\label{comparacioncampos}\end{figure}

\begin{figure}[htb]
\begin{center}
\includegraphics[width=8cm]{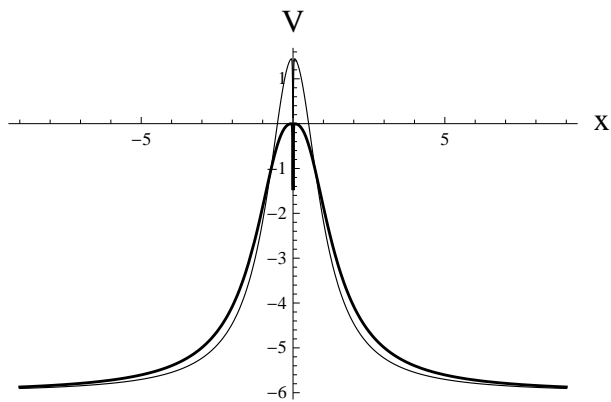}
\end{center}
\caption{Comparison of the self-interaction potentials.
The thin line represents the self-interaction potential for
the case IIb) and the thick line represents the self-interaction
potential for the case III ($\kappa=1$, $a_0=1$, $b=1$ and
$\varepsilon=0.01$).}
\label{comparacionpotenciales}
\end{figure}

In order to compare the cases II and III we plot $\phi^{\beta}(x)$
and the self-interaction potentials for both cases in Figs.
\ref{comparacioncampos} and \ref{comparacionpotenciales},
respectively. Although there are some small differences for these
quantities near the origin of the extra coordinate, similarly to the
results obtained in \cite{farakos1}, one can see that there is no
significant difference between both cases.

As we have already mentioned, in all the analyzed cases the
consistency conditions (\ref{ligaduras}) are satisfied. Since the
couplings $L(\phi)$ are bounded, the corresponding zero tensor modes
are localized on the brane. Therefore, in our model (\ref{action})
there are field configurations that give rise to a regular and
asymptotically $AdS_5$ geometry with a 4D massless spin 2 field
localized on the brane. Unlike the zero tensor modes, it is much
more difficult to analytically find the full massive spectrum. In
\cite{andrianov} it was shown that in the case II, where the
Gauss-Bonnet term is absent, the massive spectrum is continuous for
the class of solutions (\ref{asint}). In other words, for the case
II we have a localized zero tensor mode and a tower of continuous
massive modes without a mass gap. It is more difficult to
characterize the massive spectrum for cases I and II because the
equation for the mass spectrum does not have the form of a
Schr\"{o}dinger equation, at least in terms of the coordinate
$x=bw$; thus, these cases will not be considered here, but they will
be treated in a later work.

\section{Conclusions}

We have explored a thick braneworld model where the matter scalar
field is coupled non-minimally to the Einstein-Hilbert term, in
addition to this, there is a Gauss-Bonnet term in the bulk. We
compute the 4D Planck mass in terms of the 5D Planck mass and
quantities related both to the matter and the geometry. In contrast
to theories  where the  matter field is minimally coupled to the
Einstein-Hilbert term, in our model $M_{\rm Pl}$ depends explicitly
on the matter content of the bulk due to the non-minimal coupling of
the scalar field to gravity. In addition to this, by imposing
certain natural conditions on the parameters of the model (see
(\ref{ligaduras})), the expression for the 4D Planck mass (see
(\ref{masaplanck})) tells us that if a non-minimally coupled scalar
field and/or a bulk Gauss-Bonnet term are/is considered, the
predicted 4D Planck mass will be smaller than that resulting in a
model where the scalar field is minimally coupled to gravity
and the Gauss-Bonnet term is absent.

On the other hand, a relation among smoothness of geometry,
finiteness of the 4D Planck mass and localization of the tensorial
modes was studied for a wide class of solutions. Our results show
that if the geometry is regular, a finite 4D Planck mass implies
localization of gravity on the brane. In the general analysis
described above some assumptions were made about the regularity of
the geometry and its asymptotic behaviour.

We further explored an example in which we applied this general
analysis to a regular, asymptotically $AdS_5$ geometry. Since the 4D
Planck mass and the normalization condition explicitly depend on the
profile of the scalar field, it is important to solve the equation
for $\varphi$ and then to check carefully whether the conditions
(\ref{ligaduras}) are fully satisfied, since in such quantities
there are terms with negative sign contributions. We perturbatively
solved the equation for $\varphi$ by considering a small parameter
defined in terms of the strength of the non-minimal coupling. In
order to solve the latter equation it was necessary to apply the
singular perturbation method since the scalar field appeared to have
different scales of variation on different regions, a situation
known in the literature as the boundary layer problem.
The application of such a method is a difficult
task in general, since the involved differential equations are hard
to solve analytically in exact or approximate form. However, for a
wide class of solutions we managed to get perturbative analytic
expressions for $\varphi$ which satisfy all the physically
meaningful conditions (\ref{ligaduras}).

In order to elucidate the physical effects of the inclusion of the
non-minimal coupling and of the Gauss-Bonnet term, we further
considered three cases in which we switched on/off the corresponding
parameters. When the model has minimal coupling with a Gauss-Bonnet
term, the solution can be  obtained exactly and $\varphi$ has the
form of a kink/anti-kink. In the second case, when the non-minimal
coupling is considered and the Gauss-Bonnet term is switched off,
there was obtained a class of perturbative solutions with kink-like
behaviour.

For this latter case we can observe an interesting
effect: there are some configurations where the self-interaction potential of the scalar field
approaches different asymptotic values at $-\infty$ and $+\infty$,
mimicking two distinct cosmological constants at both ends of the
extra coordinate. Since the space-time is asymptotically $AdS_5$,
this means that the non-minimal coupling of the scalar field to
gravity also contributes to the ``total" cosmological constants at
$-\infty$ and $+\infty$, compensating the unevenness of the scalar
energy distribution in order to asymptotically render an even
$AdS_5$ space-time. Usually, in General Relativity the symmetries of
the geometric background are preserved by the matter energy
distribution as a consequence of self-consistency of the Einstein
equations. We see that this is not the case anymore when one
considers a system with non-minimal coupling between the scalar
field and gravity. In particular, for a symmetric or even with
respect to the extra coordinate geometry, the ``total" cosmological
constants of the asymptotically $AdS_5$ space-time get distinct
contributions from the asymptotic values of the self-interaction
potential of the scalar field at $-\infty$ and $+\infty$, and the
non-minimal coupling between the scalar matter and gravity.

We further imposed asymptotically vanishing boundary
conditions on the field configuration of the second case, yielding a
physically meaningful solution which is even with respect to the
fifth coordinate. This solution has been compared with the solution
of the third case in which the Gauss-Bonnet term is also considered.
We found that both solutions have a very similar behaviour.

Moreover, for these two last cases we also found that the
consistency conditions (\ref{ligaduras}) impose bounds on the
perturbation parameter $\varepsilon$ and that they render a smaller
4D Planck mass compared to the case in which both the non-minimal
coupling and the Gauss-Bonnet term are absent.

It will be interesting to further consider non-minimal coupling of
the scalar field with the Gauss-Bonnet term. The investigation of
the resulting model will be the subject of forthcoming work.

\section{Acknowledgements}

This research was supported by grants CIC-4.16 and CONACYT 60060-J.
DMM and RRML acknowledge a PhD grant from CONACYT. AHA and IQ thank
SNI for support.

\end{document}